\documentclass[%
 aip,
 amsmath,amssymb,
 reprint,%
]{revtex4-1}

\usepackage{graphicx}
\usepackage{dcolumn}
\usepackage{bm}

\usepackage[utf8]{inputenc}
\usepackage[T1]{fontenc}
\usepackage{mathptmx}

\begin{document}

\preprint{AIP/123-QED}

\title[Vortex Ordering and Dynamics on Santa Fe Artificial Ice Pinning Arrays]{Vortex Ordering and Dynamics on Santa Fe Artificial Ice Pinning Arrays}

\author{Wenzhao Li}
\affiliation{Department of Physics, University of Notre Dame, Notre Dame, Indiana 46656}
\author{C. J. O. Reichhardt}
 \email{cjrx@lanl.gov}
\affiliation{ 
  Theoretical Division and Center for Nonlinear Studies, Los Alamos National Laboratory, Los Alamos, New Mexico 87544
}

\author{B. Jank{\' o}$^1$}
\author{C. Reichhardt$^2$}

\date{\today}

\begin{abstract}
We numerically examine the ordering, pinning and flow of superconducting vortices interacting with a Santa Fe artificial ice pinning array. We find that as a function of magnetic field and pinning density, a wide variety of vortex states occur, including ice rule obeying states and labyrinthine patterns. In contrast to square pinning arrays, we find no sharp peaks in the critical current due to the inherent frustration effect imposed by the Santa Fe ice geometry; however, there are some smoothed peaks when the number of vortices matches the number of pinning sites. For some fillings, the Santa Fe array exhibits stronger pinning than the square array due to the suppression of one-dimensional flow channels when the vortex motion in the Santa Fe lattice occurs through the formation of both longitudinal and transverse flow channels.  
\end{abstract}

\maketitle

\section{Introduction}
In an artificial spin ice (ASI) system \cite{Nisoli13,Skjaervo20}
the states can be effectively described as spin-like degrees
of freedom which can obey the same ice rules found for the ordering of protons 
in water ice \cite{Pauling35}
or of atomic spins in certain materials \cite{Anderson56,Bramwell01}. 
One of the first artificial spin ice systems was
constructed from coupled magnetic islands in which the magnetic moment of each
island can be described as a single classical spin  \cite{Nisoli13,Wang06}.
In this system, for
specific arrangements of the effective spins
at the vertices, the ground state obeys
the ice rules and a vertex at which four spins meet has two spins
pointing 'in' and two spins pointing 'out.'
Configurations that obey the ice rule can have long range order,
such as in square ice \cite{Nisoli13,Wang06,Morgan11,Kapaklis14},
or they can be disordered, such as in kagom{\' e} ice \cite{Nisoli13,Mengotti11}.
A wide range
of additional geometries beyond square and kagom{\' e} ice have 
been proposed \cite{Skjaervo20,Morrison13}
and realized \cite{Gilbert16a,Wang16,Drisko17,Perrin16,Farhan19},
including mixed geometries which force the formation of excited
vertices \cite{Morrison13,Gilbert16a,Nisoli17}.

In addition to  studies in 
magnetic ASI systems, there are several particle-based realizations of
ASI \cite{OrtizAmbriz19}, where
a collection of interacting particles
is coupled to some form of substrate to create states
that obey the ice rules. Such systems have been studied for
colloidal particles coupled to ordered substrates
\cite{Libal06,OrtizAmbriz16,Lee18,Libal18a,Libal18},
magnetic skyrmions \cite{Ma16}, 
and vortices in type-II superconductors with nanostructured pinning arrays
\cite{Libal09,Latimer13,Trastoy14,Ge17,Wang18,Chen20,Lyu20}. 
The ice rule obeying states often arise through different mechanisms
in the particle-based systems compared to
magnetic ice systems,
since the particle ices minimize
Coulomb energy rather than vertex energy \cite{Libal18a,Nisoli18}. 
In this work we consider vortices
interacting with a variation of the ASI geometry which is called
a Santa Fe spin ice.     

Santa Fe (SF) spin ice, shown in Fig.~\ref{fig:1}(a), can contain both frustrated and
non-frustrated vertices \cite{Morrison13,Nisoli17,Zhang20},
forcing some vertices to be in an
excited state.
In the particle-based model, this would mean that some fraction of
the particles are close together, whereas the ground state of
a square ice array does not include such close particle spacings.
In superconducting ASI systems with non-frustrated ground states, 
a series of peaks appear in the critical current
at the fields corresponding to ice rule obeying states as 
well as at higher matching fields \cite{Latimer13,Trastoy14,Wang18}.

In this paper, we use numerical simulations to investigate
the vortex configurations, pinning, and flow patterns
in a system with a Santa Fe 
ASI geometry. We find numerous distinct vortex
patterns for increasing magnetic field or different pinning densities.
At the the half matching field,
the vortex configurations are close to those
expected for the ground state of the SF ice.
Compared to the square pinning lattice, in the SF ice
we find only weak or smeared
peaks in the critical depinning current, and
we show that the vortex flow patterns
are much more disordered.
For dense pinning arrays,
the ice rule obeying states vanish but the
vortices form a labyrinthine pattern.  

\section{Simulation}
We consider a two-dimensional system
of $N_{v}$ vortices interacting
with an ordered array of pinning sites
which are placed either in a Santa Fe artificial spin ice pattern
or in a square lattice.
The system contains $N_{p}$ pinning sites.
The number of vortices is proportional to the
applied magnetic field $B$, and $N_v=N_p$ corresponds to the matching
condition $B/B_{\phi}=1.0$, where $B_{\phi}$ is the matching field.
There are periodic boundary conditions in the $x$ and $y$-directions and the
equation of motion for a vortex $i$ is given by
\begin{equation} 
\eta \frac{d {\bf R}}{dt} = {\bf F}^{vv}_{i} + {\bf F}^{p}_{i} + {\bf F}^{d} \ .
\end{equation}
The damping constant $\eta$ is set to unity.
The vortex-vortex interactions are repulsive and given by
${\bf F}^{vv}_{i} = \sum F_{0}K_{1}(R_{ij}/\lambda){\hat {\bf R}}_{ij}$,
where $K_{1}$ is the modified Bessel function,
$R_{ij}$ is the distance between vortex $i$ and vortex $j$,
and $F_{0} = \phi^{2}_{0}/2\pi\mu_{0}\lambda^3$.
We set the penetration depth to $\lambda=1.8$.
In the absence of pinning sites, the
vortices form a triangular lattice.
A uniform driving force ${\bf F}^{d}=F^d{\bf \hat x}$ is applied
to all the vortices,
and the system is considered depinned when the average
steady-state vortex velocity is larger than a non-trivial value.

\begin{figure}
\includegraphics[width=\columnwidth]{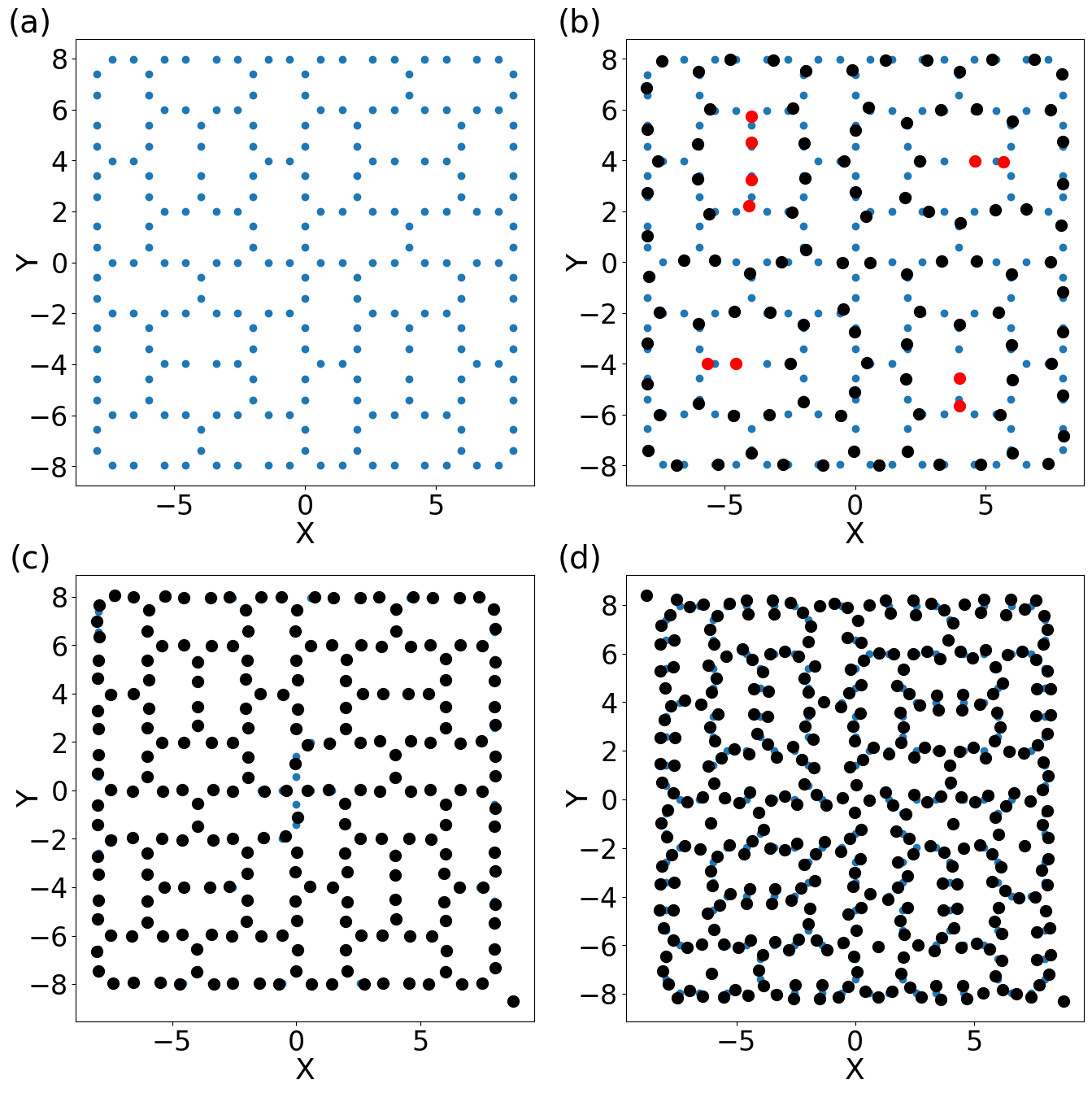}
\caption{\label{fig:1}
(a) Blue dots indicate the pinning sites arranged in a Santa Fe ASI geometry
with $d=0.825$. 
(b) The vortex positions (black dots) and pinning locations (blue dots)
for the system in panel (a) at 
$B/B_{\phi} = 1/2$, where the ice rule is mostly obeyed but
there are scattered excitations present (red dots).
(c) $B/B_{\phi} = 1.0$. (d) $B/B_{\phi} = 1.5$.  }
\end{figure}

The pinning sites are modeled as localized
traps of radius $r_{p}$ with the form  
$F^{p}_{i} = -\sum^{N_{p}}_{1}F_{p}R_{ik}\exp(-R^2_{ik}/r^2_{p}){\hat {\bf R}}_{ik}$
where $R_{ik}$ is the distance between vortex $i$ and pin $k$, and we set
$r_{p} = 0.6$.
We match our system geometry to the experiments on square
vortex ice systems \cite{Latimer13}, where $d$ is the distance
between two pinning sites.  
In Fig.~\ref{fig:1}(a) we show an
example of a Santa Fe pinning array containing four cells.
Each cell is divided into eight elementary rectangular plaquettes
in which the pinning sites can be grouped into pairs
that are spaced a distance $d=0.825$ apart.
The vortex configurations are obtained by simulated annealing from 
a high temperature.

\section{Results}
In Fig.~\ref{fig:1}(b) we show the
vortex configurations in the SF geometry for
a system with $d = 0.825$
at $B/B_{\phi} = 1/2$.
Since the vortices
are repulsive, they move as far away from each other as possible;
however, when pinning
is present, there is a competing pinning energy that favors having
the vortices occupy the pinning sites.
At $B/B_{\phi}= 1/2$, two neighboring  
pinning sites can be regarded as a single double well trap.
An individual vortex can occupy one end of this double trap,
determining the direction of the effective spin.
The lowest 
energy per vertex would have two effective spins pointing ``out'' (away from the
vertex) and two effective spins pointing ``in'' (toward the vertex);
however, due to the geometric constraints,
there must be some vertices with
two spins in and one spin out,
giving an energy higher than the ground state.
A single cell in the SF ice contains
four rectangular plaquettes that surround an inner square.
In Fig.~\ref{fig:1}(b), most of the vortices in the rectangular regions 
can form the
two-out, two-in
ground state; however, in the inner square there are
several locations where two vortices are close together
in neighboring pins, creating a high energy excitation.
The overall 
configuration is close to the expected
ice rule obeying state with forced excitations,
as predicted for the magnetic version of the 
SF spin ice \cite{Morrison13,Zhang20}.
In Fig.~\ref{fig:1}(c) we illustrate the 
vortex configurations for $B/B_{\phi} = 1.0$ at the commensurate field.
Figure~\ref{fig:1}(d) shows the configurations at
$B/B_{\phi} = 1.5$, where there are numerous
instances of individual pinning sites capturing two vortices
to form a vortex dimer state, along with
several cases where vortices are located in
the interstitial regions
in the middle of the rectangular plaquettes.
In this case, there is no long range order. For 
$B/B_{\phi} = 2.0$ and $2.5$ (not shown), the system remains disordered.

\begin{figure}
\includegraphics[width=\columnwidth]{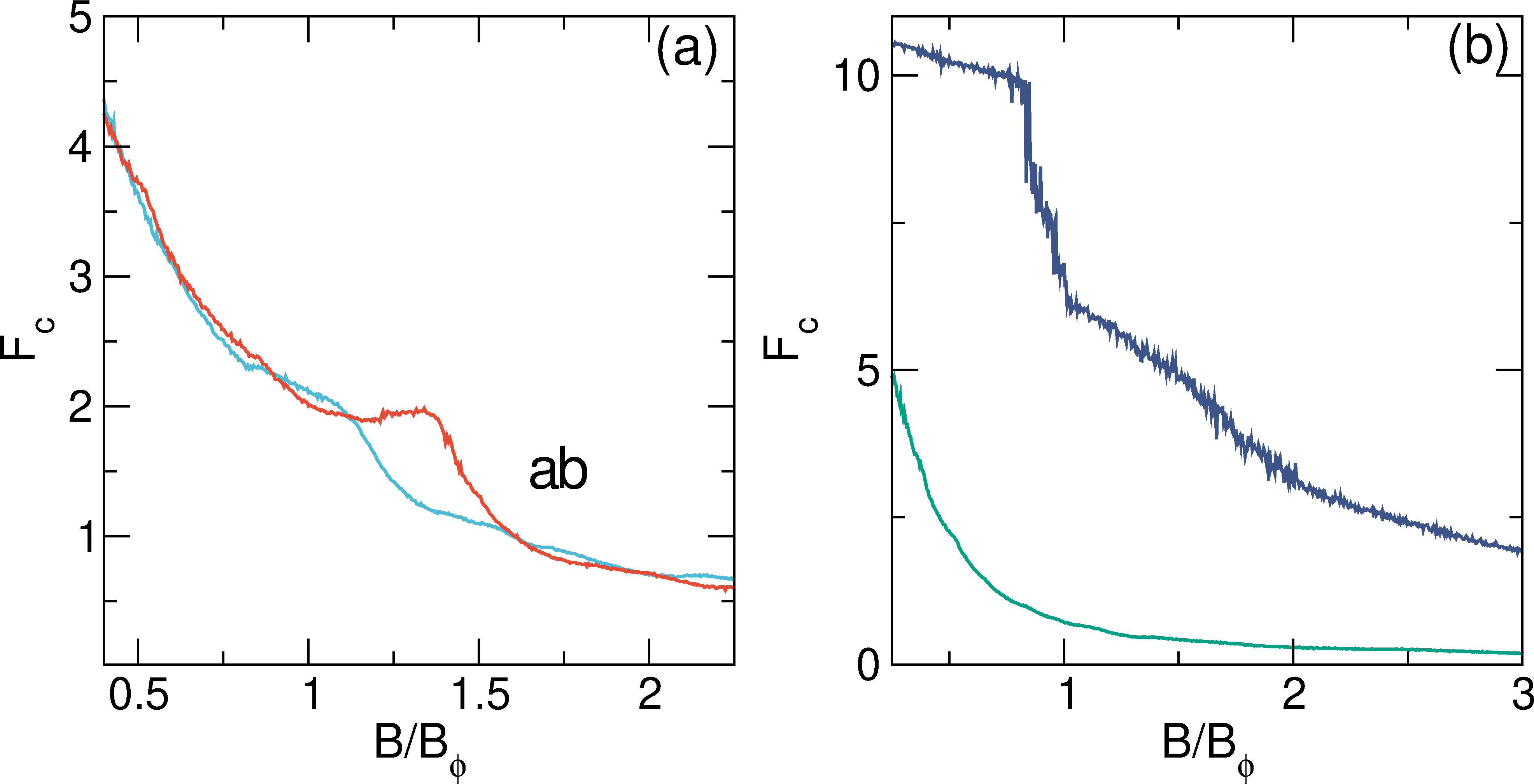}
\caption{\label{fig:2}
(a) The critical depinning force $F_c$ vs $B/B_{\phi}$ 
for the SF system in Fig.~\ref{fig:1}(a) (blue curve)
with $d=0.825$ and
for a square pinning array (red curve).
The matching field $B_\phi$ is for the SF array;
in these units, the matching field
for the square array is at
$1.286 B_{\phi}$.
The labels a and b indicate the value of $B/B_{\phi}$ at which
the images in Fig.~\ref{fig:3} were obtained.
(b) The critical depinning force $F_c$ vs $B/B_{\phi}$
for SF systems with different densities
of $d = 0.4$ (green curve) and $d=1.8$ (blue curve).
}
\end{figure}

\begin{figure}
\includegraphics[width=\columnwidth]{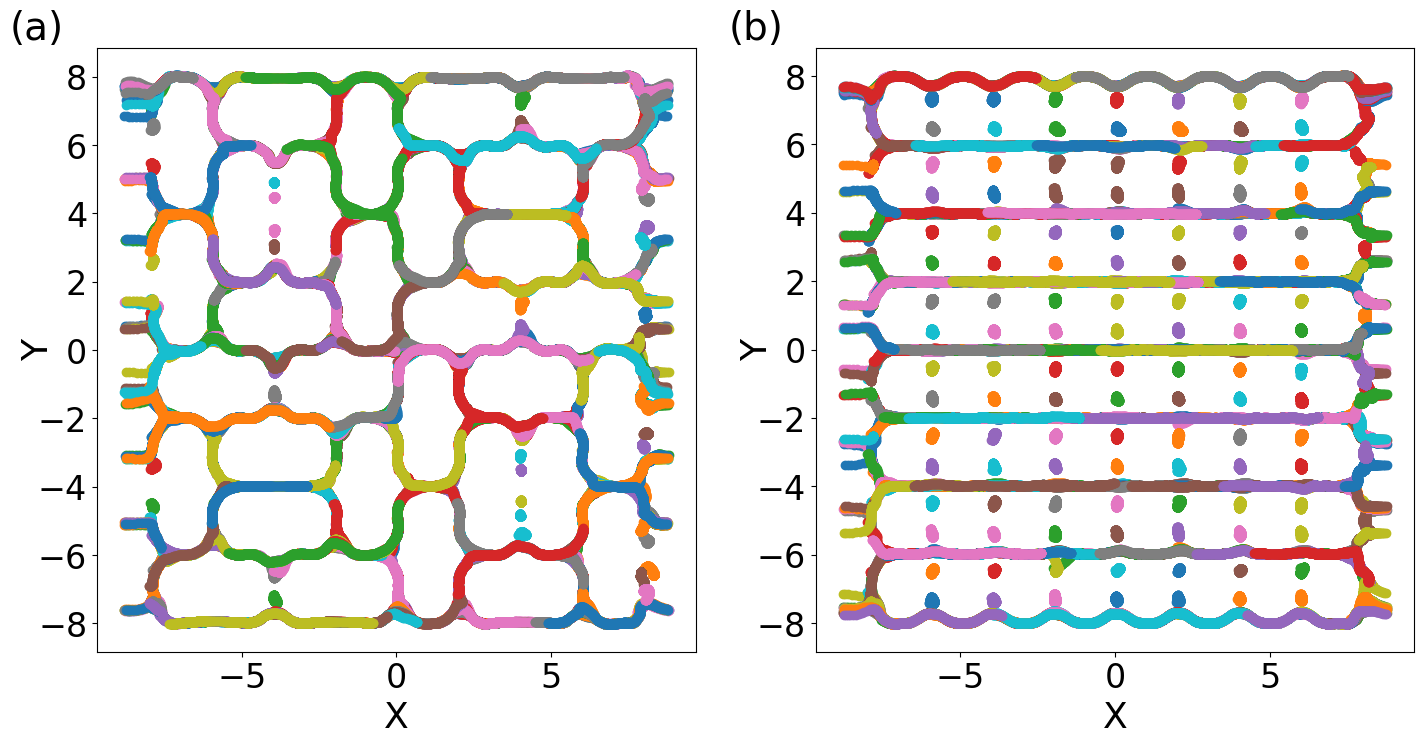}
\caption{\label{fig:3}The vortex trajectories for the system
in Fig.~\ref{fig:2}(a) with $d=0.825$ at $B/B_{\phi} = 1.67$.
(a) The SF lattice exhibits winding labyrinthine flow channels.
(b) The square lattice has easy flow one-dimensional
channels. The different colors correspond to different times.  
}
\end{figure}

In Fig.~\ref{fig:2}(a) we plot the critical depinning force
$F_c$ which is proportional to the critical current
as a function of $B/B_{\phi}$ for the SF system in Fig.~\ref{fig:1}
and for a square pinning lattice.
The $x$ axis is normalized by the matching field $B_\phi$ for the 
SF lattice, and in these units the matching field of the square lattice
falls at
$1.286 B_{\phi}$. 
The square lattice exhibits a pronounced peak
in $F_c$ at the matching field similar to that found in other studies of
vortex pinning on square substrates \cite{Baert95,Reichhardt96a}.
In the SF lattice, there is instead a broadened peak
in $F_c$ around the 
first matching field.
For a square ice system at $B/B_\phi = 1/2$,
previous work \cite{Latimer13} has shown that a
peak corresponding to the ice rule obeying ground state
appears in the critical current 
that is as large as the matching peak at $B/B_{\phi} = 1.0$.
For the SF array, there is no peak at $B/B_{\phi} = 1/2$
due to the high energy excitations that are forced
to exist by the SF geometry.
Figure~\ref{fig:2}(a) shows that the overall
pinning strength of the SF lattice is generally smaller than that of the
square lattice due the fact that there are fewer pins;
however, there are several regimes where
the depinning force for the SF geometry is higher than
that of the square array, particularly for $B/B_{\phi} > 1.0$. 
This is due to the tendency for the vortices in the
square lattice to form easy flow one-dimensional channels
along the symmetry axis of the pinning array. To 
more clearly illustrate this effect,
in Fig.~\ref{fig:3}(a,b)
we show the vortex trajectories at $B/B_{\phi} = 1.67$ for the
SF and square arrays, respectively, from Fig.~\ref{fig:2}(a)
just above depinning.
For the square array, 
the vortex motion follows one-dimensional interstitial channels
between the vortices trapped
at the pinning sites, while for the
SF array, the motion occurs through a combination of longitudinal and
transverse flow channels, 
so that some vortices move perpendicularly to the direction of drive at times.
The results in Fig.~\ref{fig:2}(a) 
also imply that for an equivalent number of pinning sites,
the SF lattice produces higher pinning than the square lattice.   

\begin{figure}
\includegraphics[width=\columnwidth]{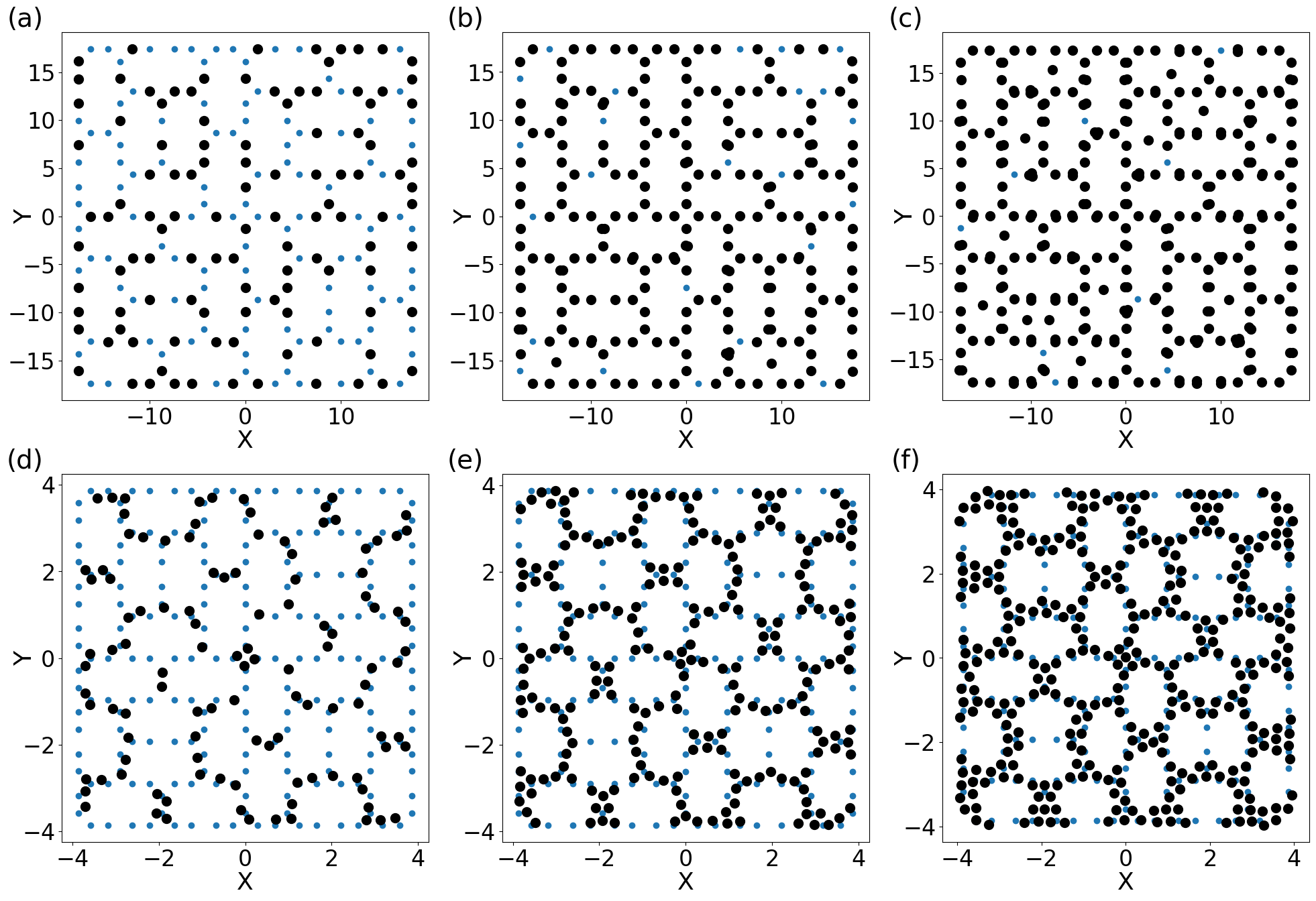}
\caption{\label{fig:4}
(a,b,c) The pinning site locations (blue dots) and vortex positions (black dots)
for the system in Fig.~\ref{fig:2}(b) at $d = 1.8$ with weak vortex
interactions
where the ice rule is lost.
(d,e,f) The same for the system in Fig.~\ref{fig:2}(b) with
$d = 0.4$, where the pinning sites begin to
overlap, creating labyrinthine vortex states.  
(a,d) $B/B_{\phi} = 0.5$.
(b,e) $B/B_{\phi}=1.0$. (c,f) $B/B_{\phi}=2.0$.
}
\end{figure}

In nanomagnetic spin ice systems, the ice rules are lost as the distance
between the magnets is 
increased due to the reduction in the pairwise interactions
between neighboring islands \cite{Wang06}.
In the superconducting 
vortex system, the effective vortex interaction can be 
tuned by changing the distance between adjacent pinning sites.
In Fig.~\ref{fig:2}(b) we plot the depinning force $F_c$ versus
$B/B_{\phi}$
for two different pinning distances,
the much larger value $d = 1.8$ which corresponds
to weak vortex interactions, and the much smaller value $d = 0.4$
which produces strong
vortex interactions.
At $d = 1.8$, the vortices are far enough apart that the pinning dominates
their dynamical behavior,
and the overall depinning threshold is much higher.
Additionally, there is no peak at $B/B_{\phi} = 1.0$, but instead
there is a downward step in the critical current.
In Fig.~\ref{fig:4}(a,b,c) we show the vortex configurations
for the $d = 1.8$ sample at $B/B_{\phi} = 0.5$, $1.5$, and
$2.0$.
For $B/B_{\phi} = 1/2$, the vertex populations are random 
and do not form ice rule obeying states.
For 
$B/B_{\phi} = 1.5$ and $2.0$,
there is a combination of doubly occupied sites and interstitial vortices.  
For a denser pinning lattice with $d = 0.4$,
there are no peaks at $B/B_{\phi}= 1/2$ or $1.0$
and the overall pinning force is reduced. 
Since the pinning radius is fixed, at the smaller $d$
the pinning sites begin to overlap, creating paths of low potential
along which the vortices can flow, thereby reducing the effectiveness of the
pinning.
This also produces increasingly labyrinthine vortex configurations,
as shown in Fig.~\ref{fig:4}(d,e,f) for $B/B_{\phi} = 1/2$, 1.5, and $2.0$.
For higher fields at this value of $d$, the labyrinthine pattern persists.
In an actual superconducting material in this regime,
the vortices could
merge to form multi-quanta states
which would have to be studied using a model different than the
point-like vortex model we consider here \cite{Tanaka02}.

\section{Summary}
We have numerically investigated vortex configurations, pinning and dynamics
in a system with a Santa Fe artificial ice pinning site arrangement. This
pinning geometry forces some vertices to occupy
higher energy states.
At half filling, the vortex configurations we observe are close to the
ground state expected for the magnetic Santa Fe ice,
with most of the vertices in low energy ice rule obeying states but
with a small number of 
high energy vertices present.
The critical depinning currents do not show a peak at the
half matching field, but exhibit a rounded peak near the fist matching field. 
For certain fillings, the Santa Fe ice
has higher pinning
than
a square pinning array array because the vortex
flow in the Santa Fe ice runs both transverse and parallel
to the driving direction, as opposed to the strictly parallel
flow that occurs in the square array.
For dense pinning arrays where the pinning sites begin to overlap,
we find that the
vortices can form an intricate labyrinthine state in the Santa Fe ice. 

\begin{acknowledgments}
We gratefully acknowledge the support of the U.S. Department of
Energy through the LANL/LDRD program for this work.
This work was supported by the US Department of Energy through
the Los Alamos National Laboratory.  Los Alamos National Laboratory is
operated by Triad National Security, LLC, for the National Nuclear Security
Administration of the U. S. Department of Energy (Contract No. 892333218NCA000001). WL and BJ were supported in part by NSF DMR-1952841.
\end{acknowledgments}

Data available on request from the authors.

\nocite{*}
\bibliography{mybib}

\end{document}